\documentclass[aps,prb,twocolumn,showpacs,preprintnumbers,amsmath,amssymb,superscriptaddress]{revtex4}%

\usepackage{graphicx}%
\usepackage{dcolumn}
\usepackage{amsmath}

\makeatletter
\def\btt#1{\texttt{\@backslashchar#1}}%
\DeclareRobustCommand\bblash{\btt{\@backslashchar}}%
\makeatother

\begin{document}


\title{Study of the magnetic penetration depth in RbOs$_2$O$_6$}

\author{R.~Khasanov}
\affiliation{ Laboratory for Neutron Scattering, ETH Z\"urich and
Paul Scherrer Institut, CH-5232 Villigen PSI, Switzerland}
\affiliation{DPMC, Universit\'e de Gen\`eve, 24 Quai
Ernest-Ansermet, 1211 Gen\`eve 4, Switzerland}
\affiliation{Physik-Institut der Universit\"{a}t Z\"{u}rich,
Winterthurerstrasse 190, CH-8057, Z\"urich, Switzerland}
\author{D.G.~Eshchenko}
\affiliation{Physik-Institut der Universit\"{a}t Z\"{u}rich,
Winterthurerstrasse 190, CH-8057, Z\"urich, Switzerland}
\affiliation{Laboratory for Muon Spin Spectroscopy, PSI, CH-5232
Villigen PSI, Switzerland}
\author{D.~Di~Castro}
\affiliation{Physik-Institut der Universit\"{a}t Z\"{u}rich,
Winterthurerstrasse 190, CH-8057, Z\"urich, Switzerland}
\affiliation{Coherentia-INFM and Department of Physics, University
of Rome "La Sapienza"}
\author{A.~Shengelaya}
\affiliation{Physik-Institut der Universit\"{a}t Z\"{u}rich,
Winterthurerstrasse 190, CH-8057, Z\"urich, Switzerland}
\author{F.~La~Mattina}
\affiliation{Physik-Institut der Universit\"{a}t Z\"{u}rich,
Winterthurerstrasse 190, CH-8057, Z\"urich, Switzerland}
\author{A.~Maisuradze}
\affiliation{Physik-Institut der Universit\"{a}t Z\"{u}rich,
Winterthurerstrasse 190, CH-8057, Z\"urich, Switzerland}
\author{C.~Baines}
\affiliation{Laboratory for Muon Spin Spectroscopy, PSI, CH-5232
Villigen PSI, Switzerland}
\author{H.~Luetkens}
\affiliation{Laboratory for Muon Spin Spectroscopy, PSI, CH-5232
Villigen PSI, Switzerland}
\author{J.~Karpinski}
\affiliation{Solid State Physics Laboratory, ETH 8093 Z\"urich,
Switzerland}
\author{S.M.~Kazakov}
\affiliation{Solid State Physics Laboratory, ETH 8093 Z\"urich,
Switzerland}
\author{H.~Keller}
\affiliation{Physik-Institut der Universit\"{a}t Z\"{u}rich,
Winterthurerstrasse 190, CH-8057, Z\"urich, Switzerland}

\begin{abstract}
Measurements of the magnetic field penetration depth $\lambda$ in
the pyrochlore superconductor RbOs$_2$O$_6$ ($T_c\simeq6.3$~K)
were
carried out by means of the muon-spin-rotation ($\mu$SR)
technique. At low temperatures $\lambda^{-2}(T)$ saturates  and
becomes constant below $T\simeq 0.2T_c$, in agreement with what is
expected for weak-coupled s-wave BCS superconductors. The value of
$\lambda$ at $T=0$ was found to be in the range of 250~nm to
300~nm. $\mu$SR and equilibrium magnetization measurements both
reveal that at low temperatures $\lambda$ is almost (at the level
of 10\%) independent of the applied magnetic field. This result
suggests
that the superconducting energy gap in RbOs$_2$O$_6$ is
isotropic.
\end{abstract}

\maketitle

\section{Introduction}

The discovery of superconductivity in pyrochlore related oxides
has attracted considerable interest in the study of these
materials.\cite{Hanawa01,Sakai01,Yonezawa04,
Hiroi04,Bruhwiller04,Kazakov04,Yonezawa04a,Yonezawa04b} However,
till now there is no agreement about the nature of superconducting
pairing mechanism in these compounds. From the one hand, based on
the results of the specific heat,\cite{Hiroi02} nuclear quadrupole
resonance (NQR) \cite{Vyaselev02} and muon-spin rotation ($\mu$SR)
\cite{Kadono04,Lumsden02} experiments, Cd$_2$Re$_2$O$_7$ is
suggested to be a weak--coupled isotropic BCS superconductor.
Specific heat,\cite{Bruhwiller04} pressure effect on the magnetic
field penetration depth \cite{Khasanov04} and nuclear magnetic
resonance (NMR) \cite{Magishi04} measurements of RbOs$_2$O$_6$ and
the band structure calculations of KOs$_2$O$_6$ \cite{Saniz04}
also point to the conventional type of superconductivity. From the
other hand,  second critical field $H_{c2}$,\cite{Hiroi04}
$\mu$SR,\cite{Kadono04,Koda04} and specific heat \cite{Hiroi04}
measurements suggest an unconventional type of paring in
KOs$_2$O$_6$ and RbOs$_2$O$_6$.

The magnetic field penetration depth $\lambda$ is one of the
fundamental lengths of a superconductor. The temperature
dependence $\lambda(T)$ reflects the quasiparticle density of
states available for thermal excitations and therefore probes the
superconducting gap structure. The shape of $\lambda(T)$ and the
zero-temperature value $\lambda(0)$ provide information about the
superconducting mechanism and
set a length scale for the
screening of an external magnetic field. In addition, the field
dependence of $\lambda$ at low temperatures may reflect the
anisotropy of the superconducting energy
gap.\cite{Kadono04,Sonier00} In this paper, we report on magnetic
field penetration depth measurements down to 30~mK in
RbOs$_2$O$_6$ by means of the transverse-field muon-spin rotation
(TF-$\mu$SR)  technique (see e.g. [\onlinecite{Zimmermann95}]).
The temperature dependence of $\lambda^{-2}$ saturates at low
temperatures and becomes constant below $T\simeq 0.2T_c$. This
behavior agrees with what is expected for weak-coupled s-wave BCS
superconductors. Measurements of the magnetic field dependence of
$\lambda$ by means of TF-$\mu$SR and magnetization reveal that at
low temperatures $\lambda$ is almost field independent. This
result
suggests
that the superconducting energy gap in
RbOs$_2$O$_6$ is isotropic. The ratio $2\Delta_0/k_BT_c$ was found
to be in the range of 3.09--3.98, which is close to the
weak-coupling BCS value 3.52.

The paper is organized as follows: In Sec.~\ref{sec:sample} we
describe the sample preparation procedure and the TF-$\mu$SR
technique as a tool to measure the magnetic field penetration
depth $\lambda$. Sec.~\ref{subsec:lambda_vs_T} comprises studies
of the temperature dependence of $\lambda$. In
Secs.~\ref{subsec:lambda_zero} and \ref{subsec:lambda_vs_H} we
discuss the calculation of the absolute value of $\lambda$ and its
magnetic field dependence. In Sec.~\ref{subsec:gap_vs_T_and_H}
results on the dependence of the zero-temperature superconducting
gap $\Delta_0$ on the superconducting critical temperature and the
magnetic field are reported. The conclusions follow in
Sec.~\ref{sec:conclusion}.

\section{Sample Preparation and Experimental
Techniques}\label{sec:sample}

\subsection{Sample preparation}\label{subsec:sample_preparation}

\begin{figure}[htb]
\includegraphics[width=1.1\linewidth]{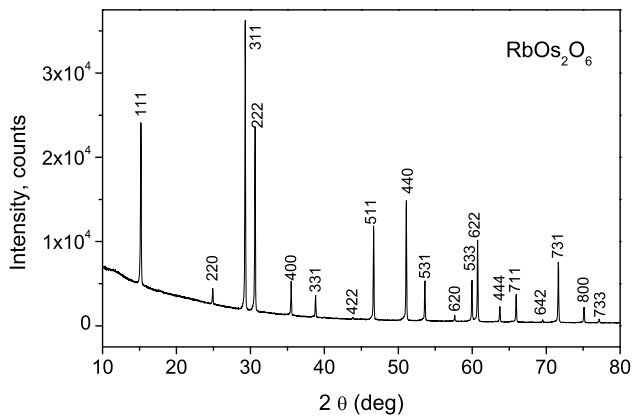}
\caption{X-ray diffraction pattern of the RbOs$_2$O$_6$ sample
synthesized in a quartz tube. All reflections are indexed on the
basis of the pyrochlore cell with a lattice parameter
a=10.1137(1)~${\rm \AA}$.}
 \label{fig:x-ray}
\end{figure}

Polycrystalline samples of RbOs$_2$O$_6$ were synthesized by a
procedure similar to that described in
Refs.~[\onlinecite{Yonezawa04,Kazakov04,Bruhwiller04}]. A
stoichiometric amount of OsO$_2$ (Alfa Aesar, 99.99\%) and Rb$_2$O
(Aldrich, 99\%) was thoroughly mixed in an argon filled dry box
and pressed into pellets. The pellets were put to a quartz tube
which was evacuated and sealed. The tube was heated up to 600$^o$C
and kept at this temperature for 24~h. According to the X-ray
analysis, the resulting sample
contained
two phases, namely,
pyrochlore RbOs$_2$O$_6$ and RbOsO$_4$. RbOsO$_4$ was removed
after 2~h etching in a 10\% solution of HCl and subsequent washing
with water and drying at 100$^o$C. The X-ray diffraction pattern
of the post treated sample is shown in Fig.~\ref{fig:x-ray} where
all reflections can be indexed on the basis of the pyrochlore cell
with a lattice parameter a=10.1137(1)~${\rm \AA}$.

\subsection{TF-$\mu$SR}\label{subsec:muSR}

The $\mu$SR experiments were performed at the $\pi$M3 beam line at
the Paul Scherrer Institute (Villigen, Switzerland). The sample
was field cooled from above $T_c$ to 30~mK in fields of 2.5~T and
1~T, and to $\simeq$1.6~K in a series of fields ranging from 5~mT
to 0.6~T.
\begin{figure}[htb]
\includegraphics[width=1.0\linewidth]{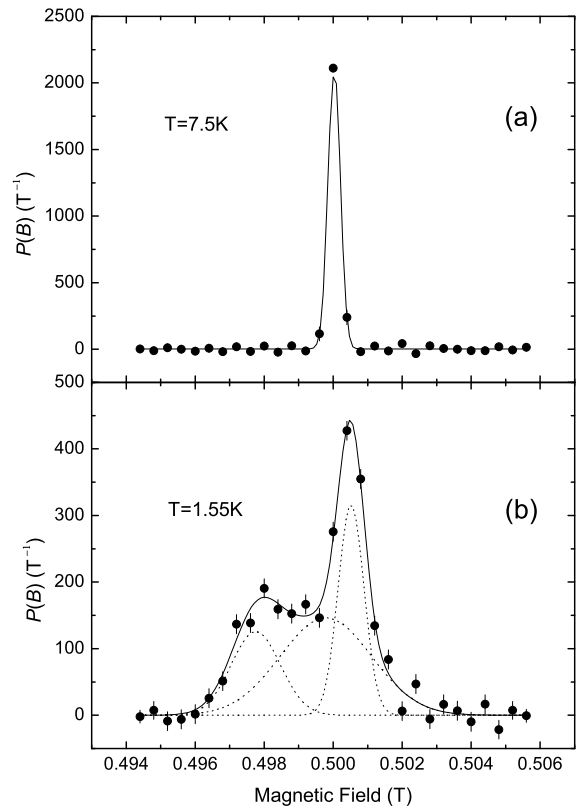}
\caption{Typical internal field distributions measured by the
$\mu$SR technique inside the RbOs$_2$O$_6$ sample above (a) and
below (b) $T_c$ after field cooling in a magnetic field of 0.5~T.
Below $T_c$ the field distribution is broadened and asymmetric.
The lines represent the best fit with Gaussian line-shapes. See
text for details.}
 \label{fig:Fourier}
\end{figure}
We used the transverse field $\mu$SR to probe the local magnetic
field distribution $P(B)$ inside the superconducting sample in the
mixed state. The second moment of $P(B)$ is connected directly
with the magnetic field penetration depth
$\lambda$.\cite{Brandt88}

The $\mu$SR signal was observed in the usual time-differential way
by monitoring the positron rate from the $\mu^+$ decay as a
function of the elapsed $\mu^+$ lifetime in the positron
telescopes. The time dependence of the positron rate is given by
the expression: \cite{msr}
\begin{equation}
 {dN \over dt} = N_0 {1\over\tau_\mu} e^{-t/\tau_\mu}
  \left[ 1 + a P(t) \right] + bg \; ,
\label{eq:N_t}
\end{equation}
where $N_0$ is a normalization constant, $bg$ is a
time-independent background, $\tau_\mu = 2.19703(4) \times
10^{-6}$~s is the $\mu^+$ lifetime, $a$ is the maximum decay
asymmetry for the particular detector telescope ($a\sim 0.18$ in
our case) and $P(t)$ is the spin polarization of the muon
ensemble:
\begin{equation}
P(t)=\int P(B)\cos(\gamma_{\mu}Bt+\phi)dB \; .
\label{eq:P_t}
\end{equation}
Here $P(B)$ is the field distribution inside a sample,
$\gamma_{\mu} = 2\pi\times135.5342$~MHz/T is the muon gyromagnetic
ratio and $\phi$ is the angle between the initial muon
polarization and the effective symmetry axis of a positron
detector. To link $P(t)$ and $P(B)$ one can use the algorithm of
Fast Fourier Transform or the direct least square fit of $P(t)$ by
the sum of precessions in discrete fields: \cite{Pomjakushin}
\begin{equation}
P(t)=\sum_i A_i \cos (\gamma_{\mu}B_i t+\phi) \; ,
\label{eq:Pom}
\end{equation}
where $A_i$ are varied and $B_i$ are fixed with spacing
$\Delta B \geq \pi /( \gamma_{\mu} t_{max})$,
$t_{max} \sim 10^{-5}$~s is the time window of the $\mu$SR
technique.

Magnetic field distributions inside the RbOs$_2$O$_6$ sample in
the normal (7.5~K) and the mixed (1.55~K) states after field
cooling in a magnetic field of 0.5~T obtained by the procedure
(\ref{eq:Pom}) are shown in Fig.~\ref{fig:Fourier}. In the normal
state, a single line at the position of the external magnetic
field with broadening arising from the nuclear magnetic moments is
seen. Below $T_c$ the field distribution is broadened and
asymmetric. For a better visualization, the fit of $P(B)$ by three
Gaussian lines is represented by dotted lines in
Fig.~\ref{fig:Fourier}. Two wide lines with the mean frequencies
below the external field are used to describe the asymmetric line
shape in the superconducting part of the sample. The narrow line
seen at a field a little bit above the external field suggests
that part of the sample is in a normal state. The superconducting
volume fraction is estimated to be $\simeq$70~\% close to the
specific heat measurements \cite{Bruhwiller04} performed on a
similarly synthesized sample where the superconducting fraction
was estimated about 80\%.

To obtain the second moment of the asymmetric field distribution
in the superconducting state we used the procedure similar to
Refs.~[\onlinecite{Grebinnik93,Weber93}]. All the $\mu$SR spectra
taken at $T < 0.85T_c$ where the three lines are resolved were
analyzed by fitting a three component expression to the $P(t)$
data:
\begin{eqnarray}
P(t)=A_b exp(-\sigma_b^2t^2/2) \cos(\gamma_{\mu}B_b t+\phi) \cr\cr
+\sum_{i=1}^2A_i exp(-\sigma_i^2t^2/2) \cos(\gamma_{\mu}B_i t+\phi) \; ,
\label{eq:gauss}
\end{eqnarray}
The first term with small $\sigma_b < 0.3$~MHz and $B_b$ close to
the applied field corresponds to the background muons stopping in
parts of the cryostat and in the nonsuperconducting parts of the
sample. The sum corresponds to the asymmetric field distribution
inside the superconductor. At $0.85T_c<T<T_c$ the two broad lines
[see e.g Fig. (\ref{fig:Fourier})] responsible for superconducting
state merge each other and the analysis is statistically correct
for one superconducting signal. At $T>T_c$ the analysis is
simplified to the background term only with $\sigma_b =\sigma_{nm}
\sim 0.1$~MHz resulting from the nuclear moments of the sample.

The superconductinq term in Eq.~(\ref{eq:gauss}) is equivalent to
the field distribution:
\begin{equation}
P(B)=\gamma_{\mu}\sum_{i=1}^2{A_i \over \sigma_i}
exp\left(-{\gamma_{\mu}^2(B-B_i)^2 \over 2\sigma_i^2}\right) \; .
\label{eq:P_B}
\end{equation}
For this distribution the mean field and the second moment are
\cite{Grebinnik93,Weber93}
\begin{equation}
\langle B \rangle=\sum_{i=1}^2{A_i B_i \over A_1+A_2} \;
\label{eq:B_mean}
\end{equation}
and
\begin{equation}
\langle \Delta B^2 \rangle=\sum_{i=1}^2{A_i \over A_1+A_2} \left[
(\sigma_i/\gamma_{\mu})^2 -[B_i- \langle B \rangle]^2 \right] \; .
\label{eq:dB}
\end{equation}
The extracted second moment of the magnetic field distribution of
the vortex lattice can be expressed in frequency units
\begin{equation}
\sigma_{sc}=\left[\gamma_{\mu}^2 \langle \Delta B^2 \rangle-
\sigma_{nm}^2\right]^{1/2} \; ,
\label{eq:sigma}
\end{equation}
where $\sigma_{nm}$ is the additional broadening due to the
nuclear moments measured
at $T>T_c$.
The absolute value of $\lambda$ is obtained from the relation
\begin{equation}
 \label{eq:sigma_vs_h}
\sigma_{sc}[\mu {\rm s}^{-1}]=4.83\times 10^4 (1-h)
  [1+3.9(1-h)^2]^{1/2}
 \lambda^{-2}[{\rm nm}] \; ,
\end{equation}
($h=H/H_{c2}$, and $H_{c2}$ is the second critical field), which
describes the filed variation in an ideal triangular vortex
lattice.\cite{Brandt88}

In separating $P(B)$ in the signal from the superconductor and
from the background by means of Gaussian functions by
Eq.~(\ref{eq:gauss}) a systematic error can occur. Part of the
background signal may in fact be associated with the
superconductor.\cite{Grebinnik93} We can estimate this error on
the assumption that the entire signal described by
Eq.~(\ref{eq:Pom}) or Eq.~(\ref{eq:gauss}) refers to the
superconductor. In this case, the second moment of the whole
$P(B)$ spectrum is systematically lower by 6.4\% at $B=0.1$~T,
5.4\% at $B=0.5$~T, 5.5\% at $B=1$~T, and 10.1\% at $B=2.5$~T.
This may result to the systematic increase of $\lambda$ by
3.2--5\% respectively.

\section{Experimental Results and Discussion}\label{seq:results_and_discussions}

\subsection{Temperature dependence of
$\lambda$}\label{subsec:lambda_vs_T}

\begin{figure}[htb]
\includegraphics[width=1.05\linewidth]{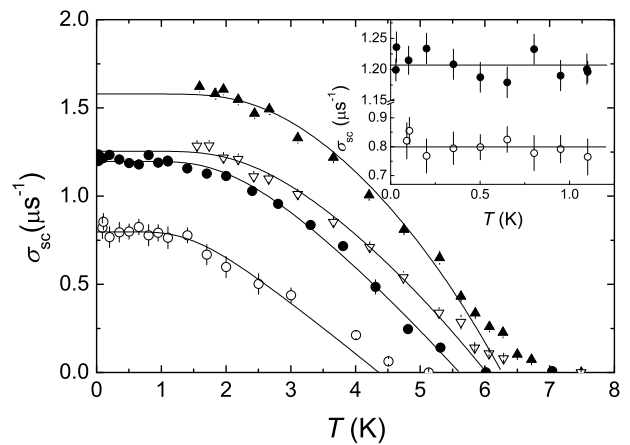}
\caption{ Temperature dependence of  $\sigma_{sc}\propto
\lambda^{-2}$ of RbOs$_2$O$_6$, measured in (from the top to the
bottom) 0.1~T, 0.5~T, 1~T, and 2.5~T fields (field-cooled).  The
inset shows the low-temperature region between 0~K and 1.25~K. The
constant (within the error bars)
$\sigma_{sc}(T)\propto\lambda^{-2}$
suggests
that RbOs$_2$O$_6$ is
a weak-coupled BCS superconductor. Lines represent fit with the
expression for the weak-coupling BCS model given in
Eq.~(\ref{eq:BCS-weak-coupled}). }
 \label{fig:lambda_vs_T}
\end{figure}

In Fig.~\ref{fig:lambda_vs_T} the temperature dependences of
$\sigma_{sc}\propto\lambda^{-2}$  for $\mu_0 H=0.1$~T, 0.5~T, 1~T,
and 2.5~T are shown.  For $\mu_0H=$1~T and 2.5~T, $\sigma_{sc}(T)$
was measured down to 30~mK. It is seen that below 1.3~K (see inset
in Fig.~\ref{fig:lambda_vs_T}) $\lambda^{-2}$ ($\sigma_{sc}$) is
{\it temperature independent}. The experimental points are well
fitted with $\sigma_{sc}(T)=const$.
Note that
the constant value of $\lambda$ at low temperatures is
predicted by the BCS model for weak-coupled
superconductors.\cite{Tinkham75} The solid lines in
Fig.~\ref{fig:lambda_vs_T} represent fit with the weak-coupling
BCS model:\cite{Tinkham75}
\begin{equation}
\frac{\lambda^{-2}(T,\Delta_0)}{\lambda^{-2}(0)}=  1+
2\int_{\Delta(T)}^{\infty}\left(\frac{\partial f}{\partial
E}\right)\frac{E}{\sqrt{E^2-\Delta(T)^2}}\  dE
 \label{eq:BCS-weak-coupled}
\end{equation}
Here, $f=[1+\exp(E/k_BT)]^{-1}$ is  the Fermi function,
$\Delta(T)=\Delta_0 \tilde{\Delta}(T/T_c)$ represents the
temperature dependence of the energy gap, $k_B$ is the Boltzman
constant, and $\Delta_0$ is the zero temperature value of the
superconducting gap. For the normalized gap
$\tilde{\Delta}(T/T_c)$ values tabulated in
Ref.~[\onlinecite{Muhlschlegel59}] were used. The data in the
Fig.~\ref{fig:lambda_vs_T} were fitted with $\sigma_{sc}(0)$ and
$\Delta_0$ as free parameters, and $T_c$ fixed from the
corresponding field-cooled magnetization ($M_{FC}$) measurements.
$T_c$ was obtained from the intersection of the linearly
extrapolated $M_{FC}(T)$ curve in the vicinity of $T_c$ with the
$M=0$ line (see inset in Fig.~\ref{fig:Hc2}). All the present
results of $\lambda(T)$ for RbOs$_2$O$_6$ are summarized in
Table~\ref{Table:lambda_results}.

\begin{table}[htb]
\caption[~]{\label{Table:lambda_results} Summary of the
$\lambda(T)$ results (see text for details).
} %
\begin{center}
\begin{tabular}{ccccccccc}\\ \hline
\hline $\mu_0H$ &$T_c$&$\Delta_0$&2$\Delta_0/k_BT_c$
&$\sigma_{sc}(0)$&\multicolumn{2}{c}{$\lambda(0)$}\\
(T)&(K)&(meV)&&($\mu {\rm s}^{-1}$)&\multicolumn{2}{c}{(nm)}\\
\hline
0.1&6.24(3)&1.07(4)&3.98(16)&1.579(11)&254(1)\footnotemark[1]&258(1)\footnotemark[2] \\
0.5&6.00(4)&0.93(3)&3.60(12)&1.254(10)&270(2)\footnotemark[1]&290(1)\footnotemark[2] \\
1&5.59(2)&0.80(3)&3.32(12)&1.197(8)&254(2)\footnotemark[1]&295(2)\footnotemark[2] \\
2.5&4.36(2)&0.58(5)&3.09(27)&0.797(8)&232(7)\footnotemark[1]&355(4)\footnotemark[2]\\

 \hline \hline \\
 \footnotetext[1]{$H_{c2}(0)$ taken from the WHH model}
 \footnotetext[2]{$H_{c2}(0)$ taken from the fit of $H_{c2}(T)$ by
means of the power law}

\end{tabular}
   \end{center}
\end{table}

\begin{figure}[htb]
\includegraphics[width=1.1\linewidth]{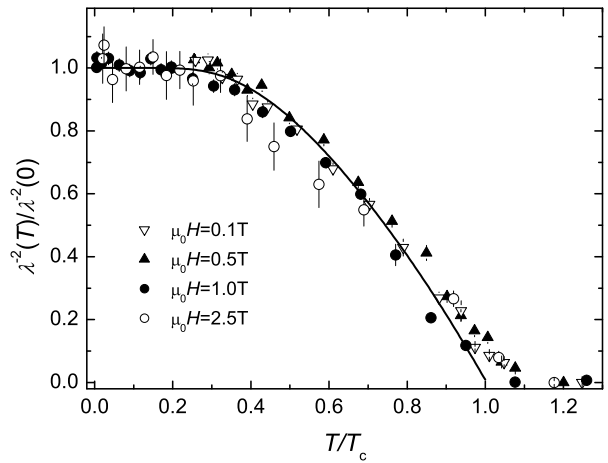}
\caption{ Normalized superfluid density
$\lambda^{-2}(T)/\lambda^{-2}(0)=\sigma_{sc}(T)/\sigma_{sc}(0)$
versus the reduced temperature $T/T_{c}$ for  0.1~T, 0.5~T, 1~T,
and 2.5~T. The solid line represents the fit of the 1~T $\mu$SR
data with Eq.~(\ref{eq:BCS-weak-coupled}). }
 \label{fig:lambda_norm}
\end{figure}

In order to compare $\lambda(T)$ obtained in different fields
(0.1~T, 0.5~T, 1~T, and 2.5~T) the normalized superfluid densities
$\lambda^{-2}(T)/\lambda^{-2}(0)=\sigma(T)_{sc}/\sigma(0)_{sc}$
versus the reduced temperature $T/T_{c}$ are plotted in
Fig.~\ref{fig:lambda_norm}. All $\lambda(T)$ collapse almost on
one curve, indicating that  the temperature dependences of
$\lambda^{-2}$ measured at different fields are nearly the same.
This is in contrast to unconventional superconductors, as e.g.
cuprate high-temperature
superconductors,\cite{Hardy93,Zimmermann95,Sonier00,Khasanovunp}
or the two-gap BCS-type MgB$_2$
superconductor,\cite{Niedermayer02} where the shape of the
temperature dependence of $\lambda^{-2}$ varies with magnetic
field. In cuprates, for example, this behavior can be explained by
the different type of symmetry of the wave function at the surface
and in the bulk ( see e.g. [\onlinecite{Muller02}]). In MgB$_2$
the field dependence of $\lambda$ is explained by the fast
suppression of the $\pi$ band by the magnetic field (see e.g.
[\onlinecite{Angst04}]). Thus, the observation of nearly the same
temperature dependences of $\lambda$  measured in different fields
is an additional argument pointing to the conventional character
of superconductivity in RbOs$_2$O$_6$.

To summarize, in the whole temperature range (down to 30~mK) the
temperature dependence of $\lambda$ is consistent with what is
expected for a weak--coupled s--wave BCS superconductor. The shape
of $\lambda(T)$ is almost independent on the magnetic field.

\subsection{The Zero Temperature Value of
$\lambda$}\label{subsec:lambda_zero}

\begin{figure}[htb]
\includegraphics[width=1.1\linewidth]{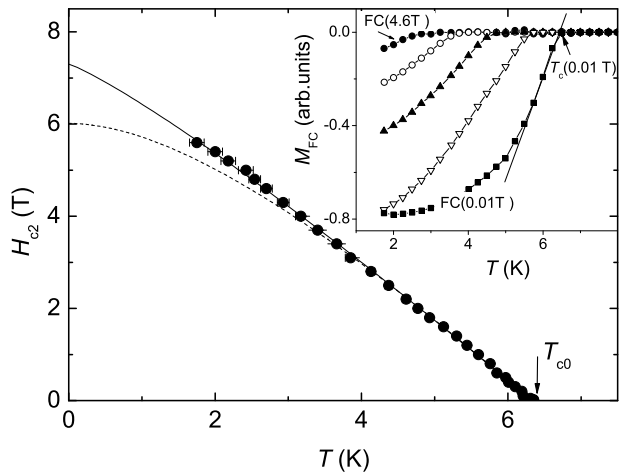}
\caption{ $H_{c2}$ vs $T$ obtained from $M_{FC}(T)$ measurements
(see text for details). The dotted line is $H_{c2}(T)$
obtained from the WHH model. The solid line is the fit with the
power law $H_{c2}(T)/H_{c2}(0)=1-(T/T_{c0})^n$ with the parameters
listed in the text. The inset shows $M_{FC}$ vs $T$ dependences
after substraction of the small paramagnetic background: from the
left to the right $\mu_0H=$ 4.6~T, 3.4~T, 2.2~T, 1~T, and 0.01~T.
}
 \label{fig:Hc2}
\end{figure}

To calculate the absolute value of $\lambda(0)$ from
$\sigma_{sc}(0)$  one needs to know the zero temperature value of
the second critical field $H_{c2}(0)$ [see
Eq.~(\ref{eq:sigma_vs_h})]. For this reason $H_{c2}(T)$ was
extracted from the $M_{FC}(T)$ curves measured in constant
magnetic fields ranging from 0.5~mT to 6~T (see
Fig.~\ref{fig:Hc2}). For each particular field $H$ the
corresponding $T_c(H)$ was taken as the temperature where
$H=H_{c2}(T=T_c)$ (see inset in Fig.~\ref{fig:Hc2}). $H_{c2}$
depends almost linearly on $T$ with some sign of saturation below
2.5~K. Note that a linear $H_{c2}(T)$ behavior was also observed
in Cd$_2$Re$_2$O$_7$ \cite{Sakai01,Jin01} and recently in
RbOs$_2$O$_6$,\cite{Bruhwiller04,Yonezawa04a} and in
KOs$_2$O$_6$\cite{Hiroi04} pyrochlore superconductors. In the
conventional BCS picture, $H_{c2}$ is linear in $T$ near $T_{c0}$
[$T_{c0}=T_c(H=0)$] and saturates by approaching 0~K. The absolute
value of $H_{c2}(0)$ can be obtained by  using the
Werthamer-Helfand-Hohenberg (WHH) formula \cite{Werthamer66}
proposed for a weak-coupling superconductor: $H_{c2}(0) =
0.693(-{\rm d}H_{c2}/{\rm d}T)| _{T=T_{c0}}$. The linear fit in
the vicinity of $T_{c0}$ yields ${\rm d}\mu_0 H_{c2}/{\rm
d}T=-1.37(4)$~T/K and $T_{c0}=6.32(19)$~K. The corresponding value
of $\mu_0 H_{c2}^{\rm WHH}(0)$ was found to be 6.00(25)~T. The
dashed line in Fig.~\ref{fig:Hc2} is the theoretical $H_{c2}(T)$
curve obtained from the WHH model in the orbital
limit.\cite{Bruhwiller04} At high temperatures (above
$\simeq$3.5~K) the WHH line agrees rather well with the
experimental data. However, at lower temperatures the experimental
points no longer follow the WHH curve, suggesting that the actual
value of $H_{c2}(0)$ is slightly larger than $H_{c2}^{\rm
WHH}(0)$.
A power law fit $H_{c2}(T)/H_{c2}(0)=1-(T/T_{c0})^n$ (solid line)
gives an exponent $n=1.17(5)$, $T_{c0}=6.33(1)$~K, and
$\mu_0H_{c2}^{PL}(0)=7.25(19)$~T.
The values of $\lambda(0)$ calculated with $H_{c2}(0)$ obtained
from the WHH model and from the fit with the power law are
summarized in Table~\ref{Table:lambda_results}. Finaly, the
representative range for $\lambda(0)$ in RbOs$_2$O$_6$ obtained
from the $\mu$SR experiments is 250-300~nm in agreement with the
low-field magnetization measurements.\cite{Khasanov04}

\subsection{The field dependence of $\lambda$
}\label{subsec:lambda_vs_H}

It is now well established that not only the temperature behavior,
but also the field dependence of $\lambda$ is completely different
for conventional BCS-type and unconventional
superconductors.\cite{Sonier00,Kadono04} By analyzing $\lambda(H)$
in different superconducting materials, it was concluded that in
superconductors associated with an anisotropic energy gap
$\lambda$ increases almost linearly with the field.\cite{Kadono04}
This effect was explained by the Doppler shift of the
quasiparticles momentum in the gap nodes.\cite{Sonier00,Kadono04}
It was also shown that in unconventional superconductors the slope
\begin{equation}
\eta=\frac{{\rm d}[\lambda(h)/\lambda(0)]}{{\rm d}h},
 \label{eq:slope}
\end{equation}
($h=H/H_{c2}$) lies in a range of 1 to 6, while it is close to
zero for superconductors with the isotropic energy
gap.\cite{Kadono04}

\begin{figure}[htb]
\includegraphics[width=1.05\linewidth]{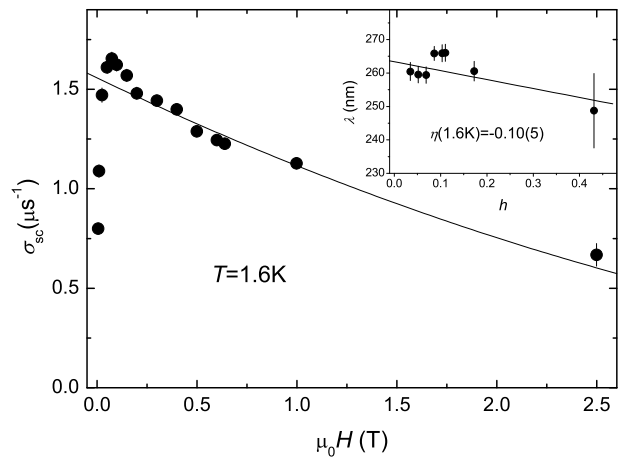}
\caption{ Magnetic field dependence of $\sigma_{sc}$ for
RbOs$_2$O$_6$ measured at $T=$1.6~K. Each point was obtained after
field-cooling the sample from a temperature above $T_c$. The solid
line is obtained from Eq.~(\ref{eq:sigma_vs_h}) with the
parameters written in the text. The inset shows $\lambda(1.6$~K)
as a function of $h=H/H_{c2}$. }
 \label{fig:lambda_vs_H_muSR}
\end{figure}

In order to obtain the field dependence of $\lambda$,
$\sigma_{sc}$ was measured as a function of the magnetic field
(see Fig.~\ref{fig:lambda_vs_H_muSR}). Each point was obtained by
field-cooling the sample from a temperature well above $T_c$ to
1.6~K. The value of $\sigma_{sc}(H,$~1.6~K) increases almost
linearly up to 75~mT; goes through a pronounced maximum around
0.1~T and then starts to decrease from 1.65~$\mu {\rm s}^{-1}$ at
the peak position to 0.67~$\mu {\rm s}^{-1}$ at 2.5~T. The solid
line is the theoretical $\sigma_{sc}(H)$ dependence obtained by
means of Eq.~(\ref{eq:sigma_vs_h}) with $\mu_0
H_{c2}(1.6$~{K})=5.80(2)~T [taken from the $H_{c2}(T)$ curve given
in Fig.~\ref{fig:Hc2}], and the field independent
$\lambda(1.6~K)$=262~nm. Above $\simeq 0.2$~T there is quite a
good agreement between theory and experimental data. The
deviations at lower fields are most probably determined by the
distortion of the vortex lattice induced by pinning. A similar
peak (followed by a plateau at high fields) is usually observed in
high-temperature superconductors.\cite{Pumpin90,Niedermayer94}

The inset in Fig.~\ref{fig:lambda_vs_H_muSR} shows the $\lambda$
vs. $h$ for $\mu_0H>0.2$~T at $T=1.6$~K. A linear fit yields
$\eta$(1.6~K)$=-0.10(5)$. The observation of a small $\eta$
suggests that RbOs$_2$O$_6$ is a superconductor with an isotropic
energy gap (see e.g. [\onlinecite{Kadono04}]).

\begin{figure}[htb]
\includegraphics[width=1.1\linewidth]{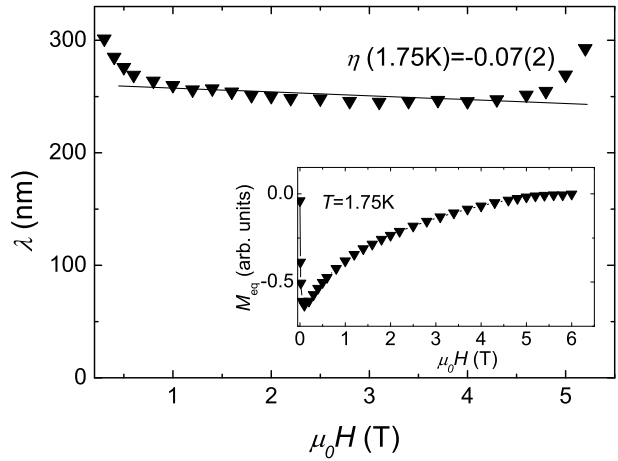}
\caption{The magnetic field dependence of  the penetration depth
$\lambda$ of RbOs$_2$O$_6$ at $T=1.75$~K, extracted from the
measurements of the equilibrium magnetization ($M_{eq}$) shown in
the inset. The solid line is the linear fit in the field region
0.6~T -- 4.8~T.}
 \label{fig:lambda_vs_H_Magn}
\end{figure}

We also performed additional $\lambda$ vs. $H$ experiments based
on measurements of the equilibrium magnetization $M_{eq}$.
Following Kogan {\it et al.} [\onlinecite{Kogan88}] one can write:
\begin{equation}
\lambda^{-2}\propto\frac{{\rm d}M_{eq}}{{\rm d}\ln H} \; .
 \label{eq:lambda_vs_H}
\end{equation}
This is the consequence of the London equation predicting that in
type-II superconductor with zero pinning, the magnetization is
proportional to $\lambda^{-2}\ln H$.  Note that
Eq.~(\ref{eq:lambda_vs_H}) is valid only in the intermediate field
region $H_{c1}\ll H\ll H_{c2}$ (here $H_{c1}$ is the first
critical field).\cite{Tinkham75} To avoid the ''pinning`` problem,
$M_{eq}(H)$  was taken from field-cooled measurements
$M_{eq}(T,H)=M_{FC}(T,H)$ (see inset in
Fig.~\ref{fig:lambda_vs_H_Magn}). As shown above in our sample
pinning (maximum in Fig.~\ref{fig:lambda_vs_H_muSR}) is suppressed
at fields above $\simeq$ 0.2~T. The $\lambda$ vs. $H$ dependence,
reconstructed by means of Eq.~(\ref{eq:lambda_vs_H}) and using the
values of $M_{eq}(H)$ at $T=1.75$~K is shown in
Fig.~\ref{fig:lambda_vs_H_Magn}. Because it is not possible to
calculate the absolute value of $\lambda$ from
Eq.~(\ref{eq:lambda_vs_H}), data in
Fig.~\ref{fig:lambda_vs_H_Magn} are scaled to the $\mu$SR value of
$\lambda(1.75$~K) at $\mu_0H=$1~T.  It is seen that between 0.6~T
and 4.8~T the $\lambda$ vs $H$ dependence is almost flat. A linear
fit of the data in this field range yields a slope
$\eta=-0.07(2)$,  in agreement with the $\eta$ value obtained from
the $\mu$SR experiment.

To summarize, the magnetic penetration depth $\lambda$ measured at
low temperatures was found to be almost (within the accuracy of
10\%) {\it field independent}. This suggests that the
superconducting energy gap in RbOs$_2$O$_6$ is {\it isotropic}.

\subsection{Dependence of the
zero-temperature superconducting gap $\Delta_0$ on the critical
temperature and the magnetic field }\label{subsec:gap_vs_T_and_H}

\begin{figure}[htb]
\includegraphics[width=1.1\linewidth]{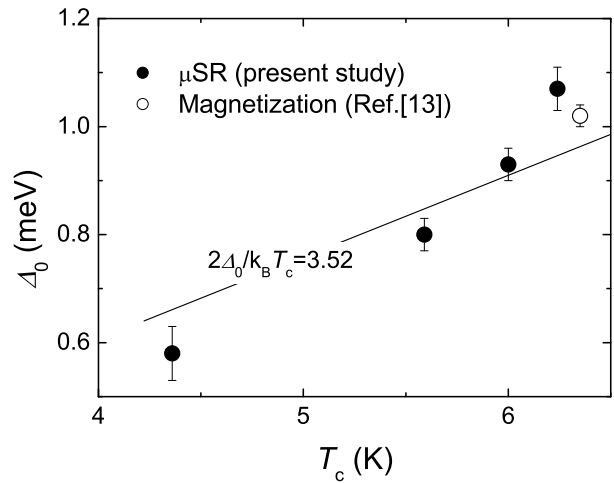}
\caption{ $\Delta_0$ vs. $T_c$ in RbOs$_2$O$_6$.  The solid line
represents the universal BCS line with 2$\Delta_0/k_BT_c=3.52$. }
 \label{fig:delta_0}
\end{figure}

\begin{figure}[htb]
\includegraphics[width=1.1\linewidth]{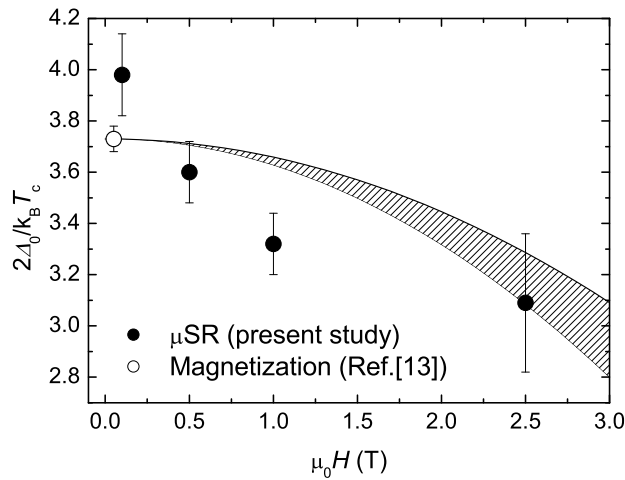}
\caption{Field dependence of  the BCS ratio $2\Delta_0/k_BT_c$.
The shadowed region represents the $2\Delta_0(H)/k_BT_c(H)$
obtained from  Eq.~(\ref{eq:Delta_vs_H}) with $H_{c2}^{\rm
WHH}(0)\leq H_{c2}(0)\leq H_{c2}^{\rm PL}(0)$. }
 \label{fig:Gap}
\end{figure}

Bearing in mind that the critical temperature $T_c$ is a function
of the applied magnetic field, in Fig.~\ref{fig:delta_0} the
zero-temperature superconducting gap $\Delta_0$
(obtained from fits
of the $\lambda^{-2}(T)$ data shown in Fig.~\ref{fig:lambda_vs_T})
are plotted as a function of $T_c$. In addition we also include in
this graph the value $\Delta_0(0.5$~mT)=1.02(2)~meV obtained from
the magnetization measurements.\cite{Khasanov04} The solid line
represents the universal BCS line with 2$\Delta_0/k_BT_c=3.52$. It
is seen that the experimental points are located close to the BCS
line. However, at high and at low temperatures the data
systematically deviate from the simple BCS line, suggesting that
the ratio $2\Delta_0/k_BT_c$ is field dependent as demonstarted in
Fig.~\ref{fig:Gap}.
It is worth noting that for conventional bulk superconductors the
ratio $2\Delta_0/k_B T_c$ is field independent, while in thin
films and in granular materials the dependence is quite
strong.\cite{Tinkham75} For the thin films of Sn, Pb and In
\cite{Morris64,Meservey64} it was experimentally observed that at
the low temperatures the magnetic field dependence of the BCS
ratio follows the empirical relation:
\begin{equation}
\frac{2\Delta_0(H)}{k_BT_c(H)} = \frac{2\Delta_0(0)}{k_BT_c(0)}
\Bigl[ 1-(H/H_{c2})^2 \Bigr],
 \label{eq:Delta_vs_H}
\end{equation}
where $2\Delta_0(0)/k_BT_c(0)$ is the BCS ratio at zero field. The
shadowed region in Fig.~\ref{fig:Gap} represents the results of
calculations by means of Eq.~\ref{eq:Delta_vs_H} with
$2\Delta_0(0)/k_BT_c (0)$=3.72
and assuming that $H_{c2}(0)$ lies between the values obtained
from the WHH model and from the fit with the power law (see
Sec.~\ref{subsec:lambda_zero}).
The field dependence of the energy gap may be explained
if
one assumes that
the electrons moving close to the surface contribute
less
to the pairing energy. \cite{Nambu63}
This may result to the
decrease of the energy gap with increasing
the
magnetic field.

To summarize, the ratio $2\Delta_0/k_BT_c$ is found in the range
of 3.09--3.98 close to the weak-coupling BCS value 3.52. The field
dependence of this ratio can be explained by the finite size of
the individual grains of the sample.

\section{Conclusions}\label{sec:conclusion}

Muon-spin rotation and magnetization studies were performed on the
pyrochlore superconductor RbOs$_2$O$_6$. The main conclusions are:
(i) The absolute value of $\lambda$ at zero temperature obtained
from $\mu$SR experiments is in the range from 250~nm to 300~nm.
(ii) In the temperature region down to 30~mK the temperature
dependence of $\lambda$ is consistent with what is expected for a
weak--coupled s--wave BCS superconductor.
(iii) The shape of $\lambda(T)$ is almost independent of the
magnetic field.
(iv) The value of the zero-temperature superconducting gap
decreases with increasing magnetic field (decreasing of $T_c$).
The ratio $2\Delta_0/k_BT_c$ was found to be in the range of
3.09--3.98 close to the weak-coupling BCS value 3.52.
(v) The $\mu$SR and the equilibrium magnetization measurements
both show that at low temperatures the magnetic penetration depth
$\lambda$ is almost (within the accuracy of 10\%) field
independent, in agreement with what is expected for a
superconductor with an isotropic energy gap.
To conclude, all the above mentioned features suggest that
RbOs$_2$O$_6$ is {\it a weak-coupled BCS superconductor with an
isotropic energy gap}.

\section{Acknowledgments}

This work was partly performed at the Swiss Muon Source (S$\mu$S),
Paul Scherrer Institute (PSI, Switzerland). The authors are
grateful to A.~Amato and D.~Herlach for providing beam-time within
the PSI short-term proposal system. This work was supported by the
Swiss National Science Foundation and by the NCCR program {\it
Materials with Novel Electronic Properties} (MaNEP) sponsored by
the Swiss National Science Foundation.

\end{document}